\documentclass[aps,prl,twocolumn,superscriptaddress,showpacs]{revtex4}
\usepackage{graphicx,graphics,amssymb,amsmath,hyperref}

\begin{document}

\newcommand{\s}{\hspace{-1pt}}

\def\etal#1{, #1}
\def\tit#1{}
\def\jour#1{#1}

\title{Giant magnetic-field dependence of the coupling between spin Tomonaga-Luttinger liquids in BaCo$_2$V$_2$O$_8$}

\author{M.~Klanj\v{s}ek}
\email{martin.klanjsek@ijs.si}
\affiliation{Jo\v{z}ef Stefan Institute, Jamova 39, and EN-FIST Centre of Excellence, Dunajska 156, 1000 Ljubljana, Slovenia}
\affiliation{Laboratoire National des Champs Magn\'etiques Intenses - CNRS, UJF, UPS and INSA, 38042 Grenoble, France}

\author{M.~Horvati\'c}
\author{S.~Kr\"amer}
\author{S.~Mukhopadhyay}
\author{H.~Mayaffre}
\author{C.~Berthier}
\affiliation{Laboratoire National des Champs Magn\'etiques Intenses - CNRS, UJF, UPS and INSA, 38042 Grenoble, France}

\author{E.~Can\'evet}
\affiliation{Institut Laue-Langevin, 38042 Grenoble, France}

\author{B.~Grenier}
\affiliation{CEA, INAC-SPSMS and Univ. Grenoble Alpes, 38054 Grenoble, France}

\author{P.~Lejay}
\affiliation{Institut N\'eel - CNRS and UJF, 38042 Grenoble, France}

\author{E.~Orignac}
\affiliation{LPENSL - CNRS, UMR 5672, 69364 Lyon, France}

\date{\today}



\begin{abstract}
We use nuclear magnetic resonance to map the complete low-temperature phase diagram of the antiferromagnetic Ising-like spin-chain system BaCo$_2$V$_2$O$_8$ as a function of the magnetic field applied along the chains. In contrast to the predicted crossover from the longitudinal incommensurate phase to the transverse antiferromagnetic phase, we find a sequence of three magnetically ordered phases between the critical fields $3.8$~T and $22.8$~T. Their origin is traced to the giant magnetic-field dependence of the total effective coupling between spin chains, extracted to vary by a factor of $24$. We explain this novel phenomenon as emerging from the combination of nontrivially coupled spin chains and incommensurate spin fluctuations in the chains treated as Tomonaga-Luttinger liquids.
\end{abstract}

\maketitle

The study of emergent phenomena in interacting quantum systems is at the heart of condensed-matter physics. Interacting fermions confined to one dimension (1D) emerge in a quantum-critical state with non-particle-like excitations, whose low-energy description is known as the Tomonaga-Luttinger liquid (TLL)~\cite{Giamarchi_2004}. As any correlation function adopts a universal form, insensitive to the microscopic details, the TLL description applies to a wide range of systems, like 1D metals~\cite{Schwartz_1998}, edge states of quantum Hall effect~\cite{Grayson_1998}, quantum wires~\cite{Jompol_2009}, carbon nanotubes~\cite{Ishii_2003} and 1D arrays of atoms on surfaces~\cite{Blumenstein_2011} or in optical traps~\cite{Vogler_2014}. The simplest and experimentally most accessible TLLs are realized in 1D quantum antiferromagnets hosting spin chains or ladders, which can be mapped onto interacting spinless fermions~\cite{Haldane_1980}. In particular, two spin-ladder systems, (C$_5$H$_{12}$N)$_2$CuBr$_4$ (BPCB)~\cite{Lorenz_2008,Klanjsek_2008,Thielemann1_2009,Thielemann2_2009,Bouillot_2011} and (C$_7$H$_{10}$N)$_2$CuBr$_4$ (DIMPY)~\cite{Hong_2010,Ninios_2012,Schmidiger_2012,Schmidiger_2013,Jeong_2013,Povarov_2014}, allowed to confirm the predicted correlation functions not only in form, but also quantitatively as a function of the magnetic field, which controls the Fermi level~\cite{Giamarchi_2004}.

While isolated TLLs cannot order because of strong quantum fluctuations, a weak coupling between TLLs leads at low temperatures to the 3D ordered state, which inherits the properties of the dominant fluctuation mode. As the Fermi surface in a TLL is reduced to two points, $k_F$ and $-k_F$, fermionic fluctuations can only occur at the wavevectors $q=0$ and $q=2k_F$~\cite{Giamarchi_2004}. In antiferromagnetic spin chains or ladders in a magnetic field, the corresponding spin fluctuations are transverse (i.e., involving spin components perpendicular to the field) {\em antiferromagnetic}, at the antiferromagnetically shifted wavevector $q=\pi$, and longitudinal (i.e., involving spin components along the field) {\em incommensurate} at the incommensurate wavevector $q=2k_F$, respectively~\cite{Giamarchi_2004}. For the Heisenberg exchange between spins, the transverse fluctuations dominate and a weakly coupled system develops a transverse antiferromagnetic order at low temperatures~\cite{Giamarchi_1999,Wessel_2000}, in the gapless region between the critical fields $B_c$ and $B_s$, which correspond to the edges of the fermion band. Examples include BPCB~\cite{Ruegg_2008,Klanjsek_2008}, DIMPY~\cite{Schmidiger_2012}, F$_5$PNN~\cite{Yoshida_2005} and copper nitrate~\cite{Willenberg_2014}. More interestingly, for the Ising-like exchange between spins and the field directed along the exchange anisotropy axis, the transverse fluctuations dominate only in the high-field region, while in the low-field region the longitudinal fluctuations dominate. Accordingly, the low-temperature ordered state in a weakly coupled system is expected to switch from the longitudinal incommensurate to the transverse antiferromagnetic as the field is increased from $B_c$ to $B_s$~\cite{Okunishi_2007}. An experimental confirmation of this important prediction reflecting the essence of the TLL description is still missing.

\begin{figure}
\includegraphics[width=1\linewidth]{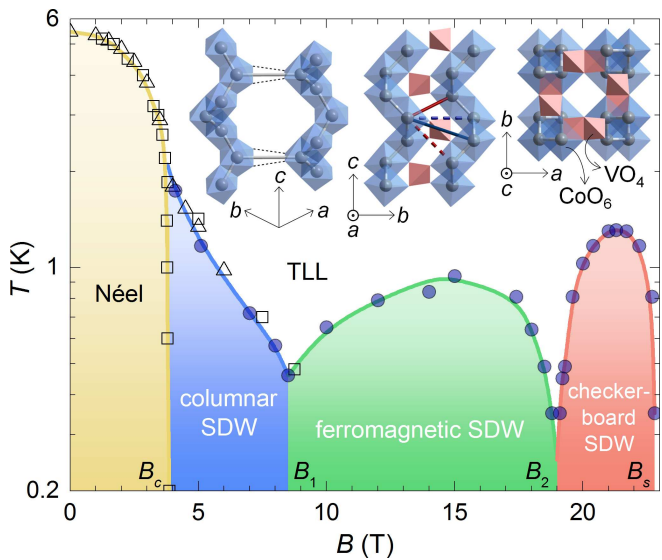}
\caption{(color online) Phase diagram of BaCo$_2$V$_2$O$_8$ as a function of temperature $T$ and the magnetic field $B$ applied along the $c$ axis. Phase boundaries are obtained from the $T_1^{-1}$ data displayed in Figs.~\ref{fig2}(a,b) (circles) and from the specific heat (squares) and neutron diffraction data (triangles) in Ref.~\cite{Canevet_2013}. Lines are guides to the eye. Inset shows three views of the structural unit of BaCo$_2$V$_2$O$_8$. Edge-sharing CoO$_6$ octahedra (blue) form chains along $c$. Exchange coupling between spin-$1/2$ Co$^{2+}$ ions (grey spheres) in neighboring chains is mediated by O$_4$ plaquetes (thin dashed lines) for NNN chains (neighbors along the $a$-$b$ bisectors) and by VO$_4$ tetrahedra (red) for NN chains (neighbors along $a$ or $b$). The intrachain coupling mediated by shorter O$_2$ bridges is much stronger. Thick lines represent different types of interchain couplings.
}
\label{fig1}
\end{figure}

Realizations of the archetypal Ising-like (i.e., $XXZ$) antiferromagnetic spin-$1/2$ chain model in the longitudinal field~\cite{Yang_1966} are therefore highly desired but unfortunately rare. Among them, BaCo$_2$V$_2$O$_8$~\cite{Wichmann_1986} has been recently studied~\cite{He1_2005,He2_2005,Kimura_2007,Kimura1_2008,Kimura2_2008,Lejay_2011,Kawasaki_2011,Zhao_2012,Ideta_2012,Canevet_2013,Kimura_2013,Niesen_2013,Niesen_2014,Grenier_2014} as one of a few with accessible critical fields, i.e., $B_c=3.8$~T and $B_s=22.8$~T~\cite{Kimura_2007}. The system develops a standard low-temperature N\'eel ordered phase in the field range up to $B_c$~\cite{He2_2005,Kimura1_2008} (Fig.~\ref{fig1}). Above $B_c$, an exotic low-temperature incommensurate phase extending up to $\mathord{\sim}9$~T was identified~\cite{Kimura1_2008,Kimura2_2008,Canevet_2013} (Fig.~\ref{fig1}), confirming the first part of the prediction~\cite{Okunishi_2007}. We continue the exploration of the phase diagram up to $B_s$ and find a sequence of {\em three} ordered phases above $B_c$, in striking contrast to the predicted crossover between the {\em two} phases. Surprisingly, all three ordered phases appear to be determined by incommensurate spin fluctuations. We show that these fluctuations combined with nontrivially coupled nearest-neighboring (NN) spin chains lead to a remarkably huge field dependence of the effective coupling between the NN chains, matching the experimental result obtained from the phase boundary. The three ordered phases then result from the competition between this variable coupling and the fixed coupling between the next-nearest-neighboring (NNN) chains, in close analogy to the frustrated spin-$1/2$ model on a square lattice~\cite{Shannon_2004}. This novel phenomenon is thus emergent in nature: although the couplings between individual spins are field-independent, the collective coupling between the 1D spin arrangements is strongly field-dependent.

\begin{figure*}
\includegraphics[width=1\linewidth]{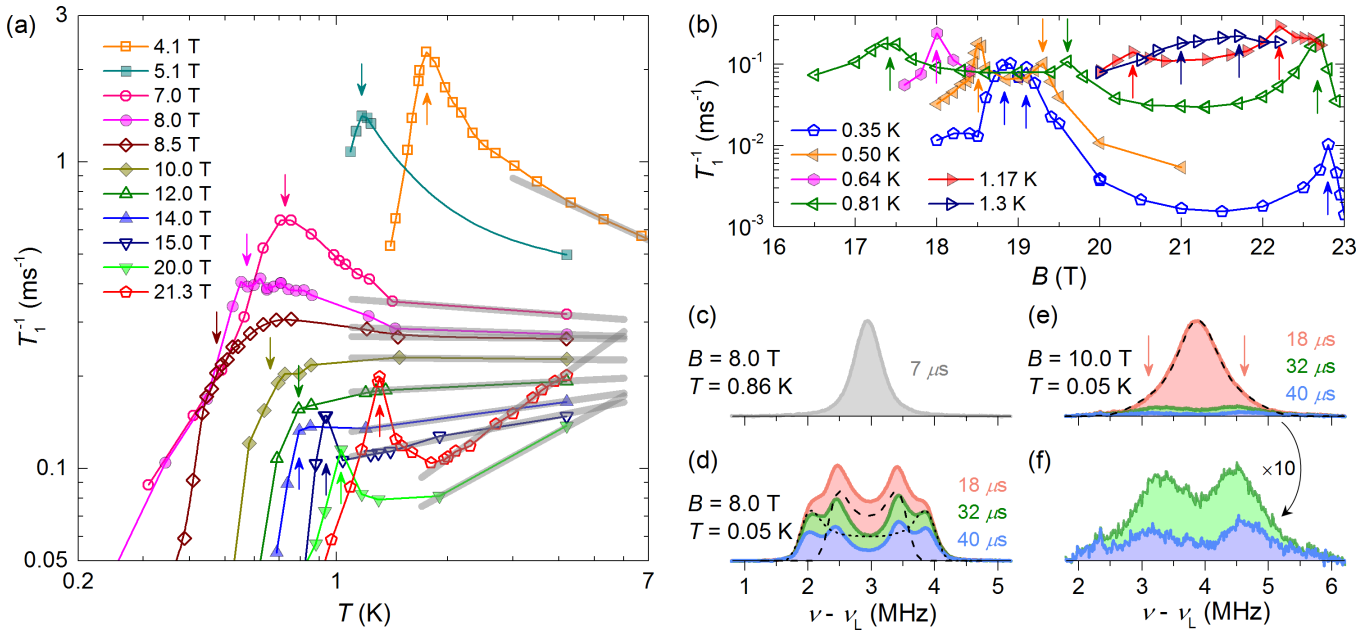}
\caption{(color online) (a) Temperature dependence of $T_1^{-1}$ in BaCo$_2$V$_2$O$_8$ for different magnetic fields $B$ applied along the $c$ axis. Thin lines are guides to the eye. Arrows mark the extracted temperatures $T_c$ of the transition from the TLL phase to the ordered phase. Thick gray lines are power-law fits in the TLL phase. (b) Field dependence of $T_1^{-1}$ for different temperatures $T$, with the presentation as in (a). (c-f) Representative $^{51}$V NMR spectra of various phases recorded after different NMR echo decay times. The frequency scale is relative to $\nu_L=\gamma B$ where $\gamma=11.193$~MHz/T and $B$ is corrected for demagnetizing effect. (c) In the TLL phase, the spectrum exhibits a single peak. (d) In the columnar SDW phase, the spectrum is reproduced by two U-shaped components of the same area. They are generated by two sets of V sites located between the NN chains hosting SDWs that are either in phase (dashed line) or in opposite phase (dotted line). (e) In the ferromagnetic SDW phase, the spectrum can be reproduced by a distribution of U-shaped components generated by a single set of V sites (dashed line). A distribution based on the 2D sinusoidal modulation of the SDW amplitude in the $a$-$b$ plane results in a spectrum with a large central weight and pronounced wings (indicated by arrows). (f) Spectra from (e) scaled up $10$-times.
}
\label{fig2}
\end{figure*}

In our $^{51}$V nuclear magnetic resonance (NMR) experiments, a single crystal of BaCo$_2$V$_2$O$_8$~\cite{Lejay_2011} was oriented with its $c$ axis, coinciding with the exchange anisotropy axis, accurately along the field direction. The phase boundary in Fig.~\ref{fig1} was mapped by means of the NMR relaxation rate $T_1^{-1}$, which directly probes the low-energy spin fluctuations~\cite{Horvatic_2002}. The temperature and field dependence of $T_1^{-1}$ measured at selected field and temperature values are shown in Figs.~\ref{fig2}(a) and (b), respectively. The datasets reveal a clear second-order phase transition through the characteristic peak indicative of the critical spin fluctuations. The transition point marks the onset of a strong $T_1^{-1}$ decrease on decreasing temperature, reflecting the suppression of spin fluctuations in the ordered phase~\cite{Beeman_1968}. In the $T_1^{-1}(T)$ datasets taken between $8$ and $14$~T, the transition peak is smeared out, most likely because of the less accurate sample orientation in the cryostat used below $14$~T, but can still be defined as the onset of the $T_1^{-1}(T)$ decrease. The overall phase boundary outlines two new ordered phases between $B_1=8.6$~T and $B_s$, and the first question is what is their nature.

Figs.~\ref{fig2}(c-f) show representative NMR spectra in various phases. The crystal structure of BaCo$_2$V$_2$O$_8$~\cite{Wichmann_1986} contains parallel chains of magnetic Co$^{2+}$ ions running along $c$ and forming a square lattice in the tetragonal $a$-$b$ plane (Fig.~\ref{fig1} inset). Above the phase boundary, throughout the TLL phase (Fig.~\ref{fig1}), the NMR spectrum exhibits a single peak [Fig.~\ref{fig2}(c)] generated by a single crystallographic V site, which is located between the NN chains. The appearance of the incommensurate order, where static spin-density waves (SDWs) form in chains, leads to the characteristic U-shaped broadening~\cite{Blinc_1981}. In the ordered phase between $B_c$ and $B_1$, the spectrum actually develops two such U-shaped components of equal area [Fig.~\ref{fig2}(d)]. This is consistent with the {\em columnar} nature of the incommensurate phase in this field range, as observed by neutron diffraction, where the symmetry between $a$ and $b$ directions is broken, so that the SDWs are in phase (in opposite phase) along one (the other) direction~\cite{Canevet_2013}. In the ordered phase above $B_1$, the NMR spectrum exhibits a single broad peak with pronounced wings. A single U-shaped component is revealed after a longer NMR echo decay time [Figs.~\ref{fig2}(e,f)], showing that the symmetry between $a$ and $b$ directions is restored. Together with the simulation of the spectral splitting~\cite{Unpublished}, this suggests a {\em ferromagnetic} nature of the incommensurate phase, meaning that all SDWs are in phase. The origin of the central narrow component observed at short echo decay times is unclear at the moment. For instance, a distribution corresponding to a long-wavelength 2D sinusoidal modulation of the SDW amplitude in the $a$-$b$ plane perfectly reproduces the lineshape [Fig.~\ref{fig2}(e)]. Alternatively, a narrow central component may also come from an eventual presence of the transverse antiferromagnetic phase coexisting with the longitudinal incommensurate phase~\cite{Unpublished}. Both scenaria require additional terms in the $XXZ$ chain Hamiltonian, which is a topic for further studies. The NMR spectrum remains qualitatively unchanged up to $B_s$ and does not offer insight into the remaining transition at $B_2=19$~T.

\begin{figure}
\includegraphics[width=1\linewidth]{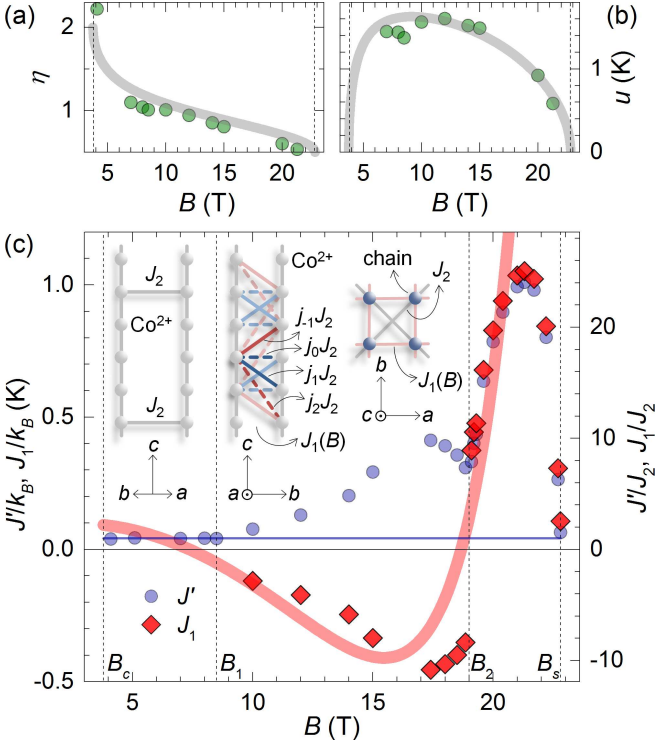}
\caption{(color online) (a,b) Dependence of the TLL parameters $\eta$ and $u$ on the magnetic field $B$ in BaCo$_2$V$_2$O$_8$ as extracted from the power-law fits in Fig.~\ref{fig2}(a). Thick gray lines are predictions of the spin-$1/2$ $XXZ$ chain model with parameters relevant to BaCo$_2$V$_2$O$_8$~\cite{Okunishi_2007}. (c) Magnetic-field dependence of the effective interchain exchange coupling $J'$ (circles) and of the exchange coupling $J_1$ (diamonds) between the NN chains as extracted from the phase boundary in Fig.~\ref{fig1}. Thin blue line is a constant fit to the $J'(B)$ data between $B_s$ and $B_1$ giving $J_2/k_B=0.042$~K. Thick red line is a fit to the $J_1(B)$ data above $B_1$ using Eq.~(\ref{eqJ1}). Inset shows the proposed schemes of interchain exchange couplings corresponding respectively to the structural units in the inset of Fig.~\ref{fig1}.
}
\label{fig3}
\end{figure}

For further insight into the nature of the ordered phases, we show that the behavior of $T_1^{-1}(T)$ above the phase boundary is quantitatively consistent with the TLL prediction for the longitudinal incommensurate fluctuations. A power-law dependence $T_1^{-1}\propto u^{-1/\eta}\,T^{1/\eta-1}$ is expected for this fluctuation mode, where $\eta$ is the interaction exponent and $u$ is the velocity of spin excitations (in kelvin units). This is obtained from the corresponding expression for the transverse antiferromagnetic fluctuations~\cite{Klanjsek_2008} by replacing $\eta$ with $1/\eta$~\cite{Okunishi_2007}. The $T_1^{-1}(T)$ dataset at $21.3$~T clearly exhibits the predicted power-law behavior, but the experimental limitations allowed us to take only a small number of data points in other datasets [Fig.~\ref{fig2}(a)]. Nevertheless, $\eta$ is determined simply from the $T_1^{-1}(T)$ slope (in a log-log scale), which exhibits a significant variation over the datasets. Moreover, once $\eta$ is known, the knowledge of the NMR form factor~\cite{Unpublished} allows us to determine $u$ from the prefactor to the power-law behavior of $T_1^{-1}(T)$ datasets, which spreads over a decade. Despite sparse datasets, we can thus estimate the field dependence of the TLL parameters $\eta$ and $u$ quite accurately. They are shown in Figs.~\ref{fig3}(a,b) together with the theoretical prediction for the spin-$1/2$ $XXZ$ chain model as applied to BaCo$_2$V$_2$O$_8$~\cite{Okunishi_2007}, and the agreement is excellent. This empirically shows a negligible sensitivity of $T_1^{-1}$ to the antiferromagnetic spin fluctuations, which would yield power-law slopes of the opposite sign, i.e., $\eta-1$ instead of $1/\eta-1$ (e.g., $-0.5$ instead of $1$ at $B_s$ where $\eta=1/2$)~\cite{Suga_2008}. Even more, as the change of spin dynamics upon transition is seen by $T_1^{-1}$ [Fig.~\ref{fig2}(a)], which is thus sensitive mainly to the incommensurate spin fluctuations, this fluctuation mode appears to condense in the ordered phase in the {\em whole} range from $B_c$ to $B_s$.

An unexpected dominance of the incommensurate mode for the nature of the ordered phases, even in the high-field region, originates from the strong enhancement of the associated total effective interchain coupling $J'$ towards $B_s$. A huge spread of the $T_1^{-1}(T)$ datasets in the TLL phase [Fig.~\ref{fig2}(a)] already {\em qualitatively} points to such an enhancement. To see how, we recall that $T_1^{-1}$ is proportional to the imaginary part of the local dynamical spin susceptibility $\chi$~\cite{Horvatic_2002}. For a system of weakly-coupled spin chains, each having the susceptibility $\chi_{\rm 1D}$, we can write $\chi=\chi_{\rm 1D}/(1-J'\chi_{\rm 1D})$ in the random-phase approximation (RPA). The $T_1^{-1}(T)$ peak at the transition point $T_c$ then translates to $J'=1/\chi_{\rm 1D}(T_c)$, where $\chi_{\rm 1D}(T_c)$ is related to $T_{1c}^{-1}=T_1^{-1}(T_c)$ in the absence of critical fluctuations, i.e., extrapolated to $T_c$ from the TLL power-law behavior. The total variation $J'(20\,{\rm T})/J'(4.1\,{\rm T})$ can then be approximated by $T_{1c}^{-1}(20\,{\rm T})/T_{1c}^{-1}(4.1\,{\rm T})\approx 20$, a surprisingly high factor. In contrast, the same approximate procedure applied to other known spin chain or ladder systems, e.g., DIMPY~\cite{Jeong_2013}, leads to an almost field-independent $J'$. We can extract $J'(B)$ in BaCo$_2$V$_2$O$_8$ also {\em quantitatively} from the phase boundary $T_c(B)$ between $B_c$ and $B_s$ using the RPA expression evaluated for the longitudinal incommensurate spin fluctuations~\cite{Okunishi_2007}:
\begin{equation}\label{eqTc}
	T_c = \frac{u}{2\pi} \left[ \frac{\Delta J'A_z}{k_B u} \sin\s\left(\s\frac{\pi}{2\eta}\s\right)
	B^2\s\s\left(\s\frac{1}{4\eta},1-\frac{1}{2\eta}\s\right) \right]^{\eta/(2\eta-1)},
\end{equation}
where $A_z(B)$ is a known amplitude of the longitudinal fluctuations~\cite{Hikihara_2004}, $\Delta$ the exchange anisotropy, $k_B$ the Boltzmann constant and $B^2(x,y)$ the square of the beta function. The resulting $J'(B)/k_B$ shown in Fig.~\ref{fig3}(c) amounts to $0.042$~K up to $B_1$, and exhibits a giant variation by a factor of $1\,{\rm K}/0.042\,{\rm K}=24$ above $B_1$.

Finally, we show that the crucial ingredient leading to the extracted giant $J'(B)$ dependence is a nontrivial, zigzag-like pattern of individual antiferromagnetic couplings between the NN chains [Fig.~\ref{fig3}(c) inset], all running along VO$_4$ tetrahedra (Fig.~\ref{fig1} inset). If we parametrize them by the weights $j_{-1}$, $j_0$, $j_1$ and $j_2$ relative to the individual antiferromagnetic coupling $J_2$ between the NNN chains [Fig.~\ref{fig3}(c) inset], the whole pattern consisting of $4$ groups of $4$ identical couplings within a helical chain period (containing $4$ chain units) sums up to
\begin{equation}\label{eqJ1}
	J_1=4J_2\Bigl[j_0+(j_{-1}+j_1)\cos{q}+j_2\cos(2q)\Bigr]
\end{equation}
in the presence of spin fluctuations with the wavevector $q$. Namely, each coupling ``tilted'' by $p$ chain units simply picks the corresponding phase shift $pq$. For incommensurate fluctuations with $q=2k_F$, the field dependence of $J_1$ comes from the relation $2k_F(B)=\pi [1-2m_z(B)]$ where $m_z(B)$ is the field-induced magnetization~\cite{Okunishi_2007}. To go further, we realize that $J_1(B)$ and $J_2$ couplings in the $a$-$b$ plane [Fig.~\ref{fig3}(c) inset] outline a pattern equivalent to the $J_1$-$J_2$ model on a square lattice~\cite{Shannon_2004}. Depending on the ratio $J_1/J_2$, this model realizes one of the three phases, each one exhibiting a different total effective coupling $J'$: columnar for $\vert J_1\vert<2J_2$ where $J'=J_2$, ferromagnetic for $J_1<-2J_2$ where $J'=-J_1-J_2$ and checkerboard for $J_1>2J_2$ where $J'=J_1-J_2$~\cite{Shannon_2004}. Applying these expressions to BaCo$_2$V$_2$O$_8$, a field-independent $J'$ in the columnar phase (between $B_c$ and $B_1$) immediately leads to $J_2/k_B=0.042$~K. In the ferromagnetic phase (between $B_1$ and $B_2$) we can then extract $J_1(B)=-J'(B)-J_2$. Finally, the phase between $B_2$ and $B_s$ should be checkerboard, hence $J_1(B)=J'(B)+J_2$. Accordingly reconstructed $J_1(B)$ dataset is shown in Fig.~\ref{fig3}(c) together with the fit using Eq.~(\ref{eqJ1}) with $j_0=8.9$, $j_{-1}+j_1=15.5$, $j_2=7.1$ and known $m_z(B)$~\cite{Kimura1_2008}. The fit fails to reproduce the $J_1(B)$ data points only close to the critical field $B_s$ where the TLL description is expected to fail anyway~\cite{Giamarchi_2004}. The obtained $J_1(B)$ reaches two orders of magnitude over $J_2$ and exhibits a remarkable sign change (to ferromagnetic), although all the individual couplings between the NN chains are antiferromagnetic. In contrast, for transverse antiferromagnetic fluctuations we set $q=\pi$ in Eq.~(\ref{eqJ1}), leading to a field-independent $J_1$, which is only of the order of $J_2$. This is the reason why the transverse antiferromagnetic order is not realized. The overall quantitative account of the $J_1(B)$ dependence a posteriori supports the previously obtained hint at the incommensurate order in the whole range between $B_c$ and $B_s$.

To conclude, our study identifies a hugely field-dependent effective magnetic coupling between spin chains in BaCo$_2$V$_2$O$_8$ as a new phenomenon emerging from the incommensurate fluctuations in TLLs coupled in a zigzag-like manner. This phenomenon could be realized also in other spin chain or ladder systems meeting the two conditions necessary for the dominance of the incommensurate fluctuations, i.e., anisotropy of the involved $S=1/2$ spins~\cite{Kimura_2007} and the zigzag-like couplings. It is important to be aware of this possibilty, as it may lead to a complicated phase diagram where a simple one would be expected. This may be the case in the recently studied BiCu$_2$PO$_6$, which satisfies both conditions~\cite{Plumb_2014,Tsirlin_2010} and exhibits a rich phase diagram~\cite{Kohama_2012}. The phenomenon could also be used deliberately as a source of tunable coupling between TLLs. For instance, implemented in an array of quantum wires, it would allow to control the current perpendicular to the wires, including its sign, simply by the gate voltage applied to the wires.

\begin{acknowledgments}
The work was partly supported by the Slovenian ARRS project No. J1-2118 and by the French ANR project NEMSICOM. The high-field part of this work was supported by EuroMagNET under the EU contract number 228043. LNCMI is a member of the European Magnetic Field Laboratory (EMFL).
\end{acknowledgments}


\begin{thebibliography}{22}

\bibitem{Giamarchi_2004}
T.~Giamarchi, {\it Quantum Physics in One Dimension} (Oxford Univ. Press, Oxford, 2004).

\bibitem{Schwartz_1998}
A.~Schwartz\etal{M.~Dressel, G.~Gr\"uner, V.~Vescoli, L.~Degiorgi, and T.~Giamarchi}, \tit{On-chain electrodynamics of metallic (TMTSF)$_2X$ salts: Observation of Tomonaga-Luttinger liquid response}\jour{Phys. Rev. B} {\bf 58}, 1261 (1998).

\bibitem{Grayson_1998}
M.~Grayson\etal{D.~C.~Tsui, L.~N.~Pfeiffer, K.~W.~West, and A.~M.~Chang}, \tit{Continuum of Chiral Luttinger Liquids at the Fractional Quantum Hall Edge}\jour{Phys. Rev. Lett.} {\bf 80}, 1062 (1998).

\bibitem{Jompol_2009}
Y.~Jompol\etal{C.~J.~B.~Ford, J.~P.~Griffiths, I.~Farrer, G.~A.~C.~Jones, D.~Anderson, D.~A.~Ritchie, T.~W.~Silk, and A.~J.~Schofield}, \tit{Probing Spin-Charge Separation in a Tomonaga-Luttinger Liquid}\jour{Science} {\bf 325}, 597 (2009).

\bibitem{Ishii_2003}
H.~Ishii\etal{H.~Kataura, H.~Shiozawa, H.~Yoshioka, H.~Otsubo, Y.~Takayama, T.~Miyahara, S.~Suzuki, Y.~Achiba, M.~Nakatake, T.~Narimura, M.~Higashiguchi, K.~Shimada, H.~Namatame, and M.~Taniguchi}, \tit{Direct observation of Tomonaga-Luttinger-liquid state in carbon nanotubes at low temperatures}\jour{Nature} {\bf 426}, 540 (2003).

\bibitem{Blumenstein_2011}
C.~Blumenstein\etal{J.~Sch\"afer, S.~Mietke, S.~Meyer, A.~Dollinger, M.~Lochner, X.~Y.~Cui, L.~Patthey, R.~Matzdorf, and R.~Claessen}, \tit{Atomically controlled quantum chains hosting a Tomonaga-Luttinger liquid}\jour{Nature Phys.} {\bf 7}, 776 (2011).

\bibitem{Vogler_2014}
A.~Vogler\etal{R.~Labouvie, G.~Barontini, S.~Eggert, V.~Guarrera, and H.~Ott}, \tit{Dimensional Phase Transition from an Array of 1D Luttinger Liquids to a 3D Bose-Einstein Condensate}\jour{Phys. Rev. Lett.} {\bf 113}, 215301 (2014).

\bibitem{Haldane_1980}
F.~D.~M.~Haldane, \tit{General Relation of Correlation Exponents and Spectral Properties of One-Dimensional Fermi Systems: Application to the Anisotropic $S=1/2$ Heisenberg Chain}\jour{Phys. Rev. Lett.} {\bf 45}, 1358 (1980).

\bibitem{Lorenz_2008}
T.~Lorenz\etal{O.~Heyer, M.~Garst, F.~Anfuso, A.~Rosch, Ch.~R\"uegg, and K.~Kr\"amer}, \tit{Diverging Thermal Expansion of the Spin-Ladder System (C$_5$H$_{12}$N)$_2$CuBr$_4$}\jour{Phys. Rev. Lett.} {\bf 100}, 067208 (2008).

\bibitem{Klanjsek_2008}
M.~Klanj\v{s}ek\etal{H.~Mayaffre, C.~Berthier, M.~Horvati\'c, B.~Chiari, O.~Piovesana, P.~Bouillot, C.~Kollath, E.~Orignac, R.~Citro, and T.~Giamarchi}, \tit{Controlling Luttinger Liquid Physics in Spin Ladders under a Magnetic Field}\jour{Phys. Rev. Lett.} {\bf 101}, 137207 (2008).

\bibitem{Thielemann1_2009}
B.~Thielemann\etal{Ch.~R\"uegg, H.~M.~R\o nnow, A.~M.~L\"auchli, J.-S.~Caux, B.~Normand, D.~Biner, K.~W.~Kr\"amer, H.-U.~G\"udel, J.~Stahn, K.~Habicht, K.~Kiefer, M.~Boehm, D.~F.~McMorrow, and J.~Mesot}, \tit{Direct Observation of Magnon Fractionalization in the Quantum Spin Ladder}\jour{Phys. Rev. Lett.} {\bf 102}, 107204 (2009).

\bibitem{Thielemann2_2009}
B.~Thielemann\etal{Ch.~R\"uegg, K.~Kiefer, H.~M.~R\o nnow, B.~Normand, P.~Bouillot, C.~Kollath, E.~Orignac, R.~Citro, T.~Giamarchi, A.~M.~L\"auchli, D.~Biner, K.~W.~Kr\"amer, F.~Wolff-Fabris, V.~S.~Zapf, M.~Jaime, J.~Stahn, N.~B.~Christensen, B.~Grenier, D.~F.~McMorrow, and J.~Mesot}, \tit{Field-controlled magnetic order in the quantum spin-ladder system (Hpip)$_2$CuBr$_4$}\jour{Phys. Rev. B} {\bf 79}, 020408 (2009).

\bibitem{Bouillot_2011}
P.~Bouillot\etal{C.~Kollath, A.~M.~L\"auchli, M.~Zvonarev, B.~Thielemann, C.~R\"uegg, E.~Orignac, R.~Citro, M.~Klanj\v{s}ek, C.~Berthier, M.~Horvati\'c, and T.~Giamarchi}, \tit{Statics and dynamics of weakly coupled antiferromagnetic spin-$1/2$ ladders in a magnetic field}\jour{Phys. Rev. B} {\bf 83}, 054407 (2011).

\bibitem{Hong_2010}
T.~Hong\etal{Y.~H.~Kim, C.~Hotta, Y.~Takano, G.~Tremelling, M.~M.~Turnbull, C.~P.~Landee, H.-J.~Kang, N.~B.~Christensen, K.~Lefmann, K.~P.~Schmidt, G.~S.~Uhrig, and C.~Broholm}, \tit{Field-Induced Tomonaga-Luttinger Liquid Phase of a Two-Leg Spin-$1/2$ Ladder with Strong Leg Interactions}\jour{Phys. Rev. Lett.} {\bf 105}, 137207 (2010).

\bibitem{Ninios_2012}
K.~Ninios\etal{T.~Hong, T.~Manabe, C.~Hotta, S.~N.~Herringer, M.~M.~Turnbull, C.~P.~Landee, Y.~Takano, and H.~B.~Chan}, \tit{Wilson Ratio of a Tomonaga-Luttinger Liquid in a Spin-$1/2$ Heisenberg Ladder}\jour{Phys. Rev. Lett.} {\bf 108}, 097201 (2012).

\bibitem{Schmidiger_2012}
D.~Schmidiger\etal{P.~Bouillot, S.~M\"uhlbauer, S.~Gvasaliya, C.~Kollath, T.~Giamarchi, and A.~Zheludev}, \tit{Spectral and Thermodynamic Properties of a Strong-Leg Quantum Spin Ladder}\jour{Phys. Rev. Lett.} {\bf 108}, 167201 (2012).

\bibitem{Schmidiger_2013}
D.~Schmidiger\etal{P.~Bouillot, T.~Guidi, R.~Bewley, C.~Kollath, T.~Giamarchi, and A.~Zheludev}, \tit{Spectrum of a Magnetized Strong-Leg Quantum Spin Ladder}\jour{Phys. Rev. Lett.} {\bf 111}, 107202 (2013).

\bibitem{Jeong_2013}
M.~Jeong\etal{H.~Mayaffre, C.~Berthier, D.~Schmidiger, A.~Zheludev, and M.~Horvati\'c}, \tit{Attractive Tomonaga-Luttinger Liquid in a Quantum Spin Ladder}\jour{Phys. Rev. Lett.} {\bf 111}, 106404 (2013).

\bibitem{Povarov_2014}
K.~Yu.~Povarov\etal{D.~Schmidiger, N.~Reynolds, A.~Zheludev, R.~Bewley}, \tit{Scaling of temporal correlations in an attractive Tomonaga-Luttinger spin liquid}\jour{arXiv:} 1406.6876 (2014).

\bibitem{Giamarchi_1999}
T.~Giamarchi and A.~M.~Tsvelik, \tit{Coupled ladders in a magnetic field}\jour{Phys. Rev. B} {\bf 59}, 11398 (1999).

\bibitem{Wessel_2000}
S.~Wessel and S.~Haas, \tit{Three-dimensional ordering in weakly coupled antiferromagnetic ladders and chains}\jour{Phys. Rev. B} {\bf 62}, 316 (2000).

\bibitem{Ruegg_2008}
C.~R\"uegg\etal{K.~Kiefer, B.~Thielemann, D.~F.~McMorrow, V.~Zapf, B.~Normand, M.~B.~Zvonarev, P.~Bouillot, C.~Kollath, T.~Giamarchi, S.~Capponi, D.~Poilblanc, D.~Biner, and K.~W.~Kr\"amer}, \tit{Thermodynamics of the Spin Luttinger Liquid in a Model Ladder Material}\jour{Phys. Rev. Lett.} {\bf 101}, 247202 (2008).

\bibitem{Yoshida_2005}
Y.~Yoshida\etal{N.~Tateiwa, M.~Mito, T.~Kawae, K.~Takeda, Y.~Hosokoshi, and K.~Inoue}, \tit{Specific Heat Study of an $S=1/2$ Alternating Heisenberg Chain System: F$_5$PNN in a Magnetic Field}\jour{Phys. Rev. Lett.} {\bf 94}, 037203 (2005).

\bibitem{Willenberg_2014}
B.~Willenberg\etal{H.~Ryll, K.~Kiefer, D.~A.~Tennant, F.~Groitl, K.~Rolfs, P.~Manuel, D.~Khalyavin, K.~C.~Rule, A.~U.~B.~Wolter, S.~S\"ullow}, \tit{Luttinger-Liquid Behavior in the Alternating Spin-Chain System Copper Nitrate}\jour{arXiv:} 1406.6149 (2014).

\bibitem{Okunishi_2007}
K.~Okunishi and T.~Suzuki, \tit{Field-induced incommensurate order for the quasi-one-dimensional $XXZ$ model in a magnetic field}\jour{Phys. Rev. B} {\bf 76}, 224411 (2007).

\bibitem{Yang_1966}
C.~N.~Yang and C.~P.~Yang, \tit{One-Dimensional Chain of Anisotropic Spin-Spin Interactions. I. Proof of Bethe's Hypothesis for Ground State in a Finite System}\jour{Phys. Rev.} {\bf 150}, 321 (1966).

\bibitem{Wichmann_1986}
R.~Wichmann and Hk.~M\"uller-Buschbaum, \tit{Neue Verbindungen mit SrNi$_2$V$_2$0$_8$-Struktur: BaCo$_2$V$_2$0$_8$ und BaMg$_2$V$_2$0$_8$}\jour{Z. anorg. allg. Chem.} {\bf 534}, 153 (1986).

\bibitem{He1_2005}
Z.~He\etal{D.~Fu, T.~Ky\^omen, T.~Taniyama, and M.~Itoh}, \tit{Crystal Growth and Magnetic Properties of BaCo$_2$V$_2$O$_8$}\jour{Chem. Mater.} {\bf 17}, 2924 (2005).

\bibitem{He2_2005}
Z.~He\etal{T.~Taniyama, T.~Ky\^omen, and M.~Itoh}, \tit{Field-induced order-disorder transition in the quasi-one-dimensional anisotropic antiferromagnet BaCo$_2$V$_2$O$_8$}\jour{Phys. Rev. B} {\bf 72}, 172403 (2005).

\bibitem{Kimura_2007}
S.~Kimura\etal{H.~Yashiro, K.~Okunishi, M.~Hagiwara, Z.~He, K.~Kindo, T.~Taniyama, and M.~Itoh}, \tit{Field-Induced Order-Disorder Transition in Antiferromagnetic BaCo$_2$V$_2$O$_8$ Driven by a Softening of Spinon Excitation}\jour{Phys. Rev. Lett.} {\bf 99}, 087602 (2007).

\bibitem{Kimura1_2008}
S.~Kimura\etal{T.~Takeuchi, K.~Okunishi, M.~Hagiwara, Z.~He, K.~Kindo, T.~Taniyama, and M.~Itoh}, \tit{Novel Ordering of an $S=1/2$ Quasi-1d Ising-Like Antiferromagnet in Magnetic Field}\jour{Phys. Rev. Lett.} {\bf 100}, 057202 (2008).

\bibitem{Kimura2_2008}
S.~Kimura\etal{M.~Matsuda, T.~Masuda, S.~Hondo, K.~Kaneko, N.~Metoki, M.~Hagiwara, T.~Takeuchi, K.~Okunishi, Z.~He, K.~Kindo, T.~Taniyama, and M.~Itoh}, \tit{Longitudinal Spin Density Wave Order in a Quasi-1D Ising-Like Quantum Antiferromagnet}\jour{Phys. Rev. Lett.} {\bf 101}, 207201 (2008).

\bibitem{Lejay_2011}
P.~Lejay\etal{E.~Canevet, S.~K.~Srivastava, B.~Grenier, M.~Klanj\v{s}ek, and C.~Berthier}, \tit{Crystal growth and magnetic property of MCo$_2$V$_2$O$_8$ (M=Sr and Ba)}\jour{J. Cryst. Growth} {\bf 317}, 128 (2011).

\bibitem{Kawasaki_2011}
Y.~Kawasaki\etal{J.~L.~Gavilano, L.~Keller, J.~Schefer, N.~B.~Christensen, A.~Amato, T.~Ohno, Y.~Kishimoto, Z.~He, Y.~Ueda, and M.~Itoh}, \tit{Magnetic structure and spin dynamics of the quasi-one-dimensional spin-chain antiferromagnet BaCo$_2$V$_2$O$_8$}\jour{Phys. Rev. B} {\bf 83}, 064421 (2011).

\bibitem{Zhao_2012}
Z.~Y.~Zhao\etal{X.~G.~Liu, Z.~He, X.~M.~Wang, C.~Fan, W.~P.~Ke, Q.~J.~Li, L.~M.~Chen, X.~Zhao, and X.~F.~Sun}, \tit{Heat transport of the quasi-one-dimensional Ising-like antiferromagnet BaCo$_2$V$_2$O$_8$ in longitudinal and transverse fields}\jour{Phys. Rev. B} {\bf 85}, 134412 (2012).

\bibitem{Ideta_2012}
Y.~Ideta\etal{Yu~Kawasaki, Y.~Kishimoto, T.~Ohno, Y.~Michihiro, Z.~He, Y.~Ueda, and M.~Itoh}, \tit{$^{51}$V NMR study of antiferromagnetic state and spin dynamics in quasi-one-dimensional BaCo$_2$V$_2$O$_8$}\jour{Phys. Rev. B} {\bf 86}, 094433 (2012).

\bibitem{Canevet_2013}
E.~Can\'evet\etal{B.~Grenier, M.~Klanj\v{s}ek, C.~Berthier, M.~Horvati\'c, V.~Simonet, and P.~Lejay}, \tit{Field-induced magnetic behavior in quasi-one-dimensional Ising-like antiferromagnet BaCo$_2$V$_2$O$_8$: A single-crystal neutron diffraction study}\jour{Phys. Rev. B} {\bf 87}, 054408 (2013).

\bibitem{Kimura_2013}
S.~Kimura\etal{K.~Okunishi, M.~Hagiwara, K.~Kindo, Z.~He, T.~Taniyama, M.~Itoh, K.~Koyama, and K.~Watanabe}, \tit{Collapse of Magnetic Order of the Quasi One-Dimensional Ising-Like Antiferromagnet BaCo$_2$V$_2$O$_8$ in Transverse Fields}\jour{J. Phys. Soc. Jpn.} {\bf 82}, 033706 (2013).

\bibitem{Niesen_2013}
S.~K.~Niesen\etal{G.~Kolland, M.~Seher, O.~Breunig, M.~Valldor, M.~Braden, B.~Grenier, and T.~Lorenz}, \tit{Magnetic phase diagrams, domain switching, and quantum phase transition of the quasi-one-dimensional Ising-like antiferromagnet BaCo$_2$V$_2$O$_8$}\jour{Phys. Rev. B} {\bf 87}, 224413 (2013).

\bibitem{Niesen_2014}
S.~K.~Niesen\etal{O.~Breunig, S.~Salm, M.~Seher, M.~Valldor, P.~Warzanowski, and T.~Lorenz}, \tit{Substitution effects on the temperature versus magnetic field phase diagrams of the quasi-one-dimensional effective Ising spin-$1/2$ chain system BaCo$_2$V$_2$O$_8$}\jour{Phys. Rev. B} {\bf 90}, 104419 (2014).

\bibitem{Grenier_2014}
B.~Grenier\etal{S.~Petit, V.~Simonet, L.-P.~Regnault, E.~Can\'evet, S.~Raymond, B.~Canals, C.~Berthier, P.~Lejay}, \tit{Longitudinal and transverse Zeeman ladders in the Ising-like chain antiferromagnet BaCo$_2$V$_2$O$_8$}\jour{arXiv:} 1407.0213 (2014).

\bibitem{Shannon_2004}
N.~Shannon\etal{B.~Schmidt, K.~Penc, and P.~Thalmeier}, \tit{Finite temperature properties and frustrated ferromagnetism in a square lattice Heisenberg model}\jour{Eur. Phys. J. B} {\bf 38}, 599 (2004).

\bibitem{Horvatic_2002}
M.~Horvati\'c and C.~Berthier, "NMR Studies of Low-Dimensional Quantum Antiferromagnets," in C.~Berthier, L.~P.~Levy, and G.~Martinez, {\it High Magnetic Fields: Applications in Condensed Matter Physics and Spectroscopy} (Springer-Verlag, Berlin, 2002), p. 200.

\bibitem{Beeman_1968}
D.~Beeman and P.~Pincus, \tit{Nuclear Spin-Lattice Relaxation in Magnetic Insulators}\jour{Phys. Rev.} {\bf 166}, 359 (1968).

\bibitem{Blinc_1981}
R.~Blinc, \tit{Magnetic resonance and relaxation in structurally incommensurate systems}\jour{Phys. Rep.} {\bf 79}, 331 (1981).

\bibitem{Unpublished}
M.~Klanj\v{s}ek\etal{} (unpublished).

\bibitem{Suga_2008}
S.~Suga, \tit{Tomonaga-Luttinger Liquid in Quasi-One-Dimensional Antiferromagnet BaCo$_2$V$_2$O$_8$ in Magnetic Fields}\jour{J. Phys. Soc. Jpn.} {\bf 77}, 074717 (2008).

\bibitem{Hikihara_2004}
T.~Hikihara and A.~Furusaki, \tit{Correlation amplitudes for the spin-$1/2$ $XXZ$ chain in a magnetic field}\jour{Phys. Rev. B} {\bf 69}, 064427 (2004).

\bibitem{Plumb_2014}
K.~W.~Plumb\etal{K.~Hwang, Y.~Qiu, L.~W.~Harriger, G.~E.~Granroth, G.~J.~Shu, F.~C.~Chou, Ch.~R\"uegg, Y.~B.~Kim, Y.-J.~Kim}, \tit{Giant Anisotropic Interactions in the Copper Based Quantum Magnet BiCu$_2$PO$_6$}\jour{arXiv:} 1408.2528 (2014).

\bibitem{Tsirlin_2010}
A.~A.~Tsirlin\etal{I.~Rousochatzakis, D.~Kasinathan, O.~Janson, R.~Nath, F.~Weickert, C.~Geibel, A.~M.~L\"auchli, and H.~Rosner}, \tit{Bridging frustrated-spin-chain and spin-ladder physics: Quasi-one-dimensional magnetism of BiCu$_2$PO$_6$}\jour{Phys. Rev. B} {\bf 82}, 144426 (2010).

\bibitem{Kohama_2012}
Y.~Kohama\etal{S.~Wang, A.~Uchida, K.~Prsa, S.~Zvyagin, Yu.~Skourski, R.~D.~McDonald, L.~Balicas, H.~M.~R\o{}nnow, Ch.~R\"uegg, and M.~Jaime}, \tit{Anisotropic Cascade of Field-Induced Phase Transitions in the Frustrated Spin-Ladder System BiCu$_2$PO$_6$}\jour{Phys. Rev. Lett.} {\bf 109}, 167204 (2012).

\end{thebibliography}
\end{document}